\documentclass[pre,twocolumn,showpacs,amsmath,amssymb]{revtex4}
\usepackage{graphicx}
\usepackage{dcolumn}
\usepackage{bm}

\begin{document}

\title{Stable smectic phase in suspensions of polydisperse colloidal platelets with identical thickness}


\author{Dazhi Sun}
\affiliation{Department of Mechanical Engineering, Texas A\&M University, College Station,
TX 77843 USA.}

\author{Hung-Jue Sue}
\affiliation{Department of Mechanical Engineering, Texas A\&M University, College Station,
TX 77843 USA.}

\author{Zhengdong Cheng}
\email{cheng@chemail.tamu.edu}
\affiliation{Artie McFerrin Department of Chemical Engineering, Texas A\& M University, College Station,
TX 77843 USA.}

\author{Yuri Mart\'{\i}nez-Rat\'on}
\email{yuri@math.uc3m.es}
\affiliation{Grupo Interdisciplinar de Sistemas Complejos (GISC),
Departamento de Matem\'{a}ticas,Escuela Polit\'{e}cnica Superior,
Universidad Carlos III de Madrid, Avenida de la Universidad 30, E--28911, Legan\'{e}s, Madrid, Spain}

\author{Enrique Velasco}
\affiliation{Departamento de F\'{\i}sica Te\'orica de la Materia Condensada
and Instituto de Ciencia de Materiales Nicol\'as Cabrera,
Universidad Aut\'onoma de Madrid, E-28049 Madrid, Spain.}

\date{\today}

\begin{abstract}
We report the nematic and smectic ordering in a new aqueous suspension of 
monolayer $\alpha$-Zirconium phosphate platelets possessing a high polydispersity in 
diameter but uniform thickness. We observe an isotropic--nematic transition as 
the platelet volume fraction increases, followed by the formation of a smectic, an elusive 
phase that has been rarely seen in discotic liquid crystals. The smectic phase is 
characterized by X-ray diffraction, high-resolution transmission electron microscopy, and 
optical microscopy. The phase equilibria in this highly polydisperse suspension is rationalized
in terms of a theoretical approach based on density--functional theory.
\end{abstract}


\maketitle

Synthetic inorganic platelets are being used extensively to investigate phase transitions in discotic 
liquid crystals consisting of colloidal, approximately disc--shaped, particles in aqueous 
suspension. Starting from the completely disordered isotropic (I) phase in the dilute regime, 
most discotic suspensions exhibit an orientationally ordered nematic (N) phase 
as the volume fraction of the platelets in the sample increases. At higher volume fractions a columnar (C) phase, with columns of 
platelets arranged in a two--dimensional lattice, has been observed in some 
materials, either directly from the I phase \cite{Brown} or via N ordering \cite{vdKooij}. 
Computer simulations on monodisperse \cite{Veerman} and polydisperse \cite{Bates} hard platelets 
of high aspect ratio show that indeed these liquid--crystalline phases can be 
stabilized solely from exclusion interactions, but no layered ordering, of smectic (S) type, is found. 
However, the work of van der Kooij et al. \cite{vdKooij} seemed to indicate that smectic or 
lamellar phases might exist in colloidal platelet suspensions at high volume fractions. Usually these 
suspensions are polydisperse in both thickness and diameter. The C phase stabilizes even for 
a considerable degree of diameter polydispersity, but uniformly thick platelets with very high diameter 
polydispersity cannot form a C phase, which presumably would be superseded by a S phase before the 
glassy or crystalline phases. Nevertheless, the S phase remains elusive \cite{vdKooij} 
and swollen lamellar phases has been observed to form, directly from 
a I suspension, only in a few materials \cite{Gabriel,Wang}.

\begin{figure}[h]
\includegraphics[width=3.0in]{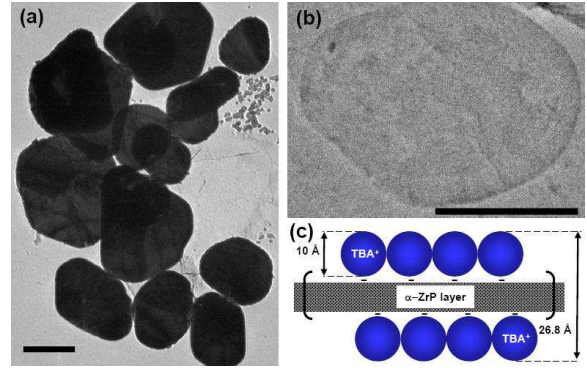}
\caption{\label{1} (Color online). $\alpha$--Zirconium phosphate platelets. (a) TEM of pristine ZrP platelets.
(b) TEM of a ZrP monolayer platelet exfoliated by TBA. (c) Thickness schematic of a ZrP monolayer
platelet. Length is not drawn to scale. The scale bars in (a) and (b) are 1 $\mu$m.}
\end{figure}
Here we report the observation of a stable S phase in a new system of platelets with high 
polydispersity in diameter but identical thickness. 
Contrary to the approximately hard--core platelets analyzed in Ref. \cite{vdKooij}, our particles
interact via long--range forces. We find that, as the 
platelet volume fraction $\phi$ increases, the samples follow the phase sequence I--N--S, with no stable C phase 
in the volume--fraction interval explored. Some features of the phase transitions can be explained in 
terms of a density--functional theory (DFT) for a simple interaction model.

Pristine $\alpha$-Zirconium phosphate (ZrP) layered platelets were synthesized 
through a hydrothermal method similar to 
that reported in Ref. \cite{report} and exfoliated with tetra-n-butylammonium hydroxide 
(TBA$^+$OH$^-$, Aldrich, 1 mol/l in methanol) at a molar ratio of ZrP:TBA = 1:1 in de-ionized water.
No salts were added during the experiments. After reaction, the platelets were allowed to expose in 
sonication for 5 mins and then sit for one day to reach phase transition equilibrium.
The platelets, which contain two TBA monolayers, have an identical thickness of $L=2.68$ nm.
This process has been demonstrated \cite{report} to lead to
stable aqueous colloidal suspensions (see Fig. \ref{1}). The average platelet is cylindrical but 
with relatively high variations in
transverse shape. Samples containing various platelet volume fractions but the same
polydispersity were prepared. Transmission Electron Microscopy (TEM) images [Fig. \ref{1}(b)] allow for 
an approximate estimation of mean particle diameter ($\sim 2$ $\mu$m).

\begin{figure}[h]
\includegraphics[width=3.0in]{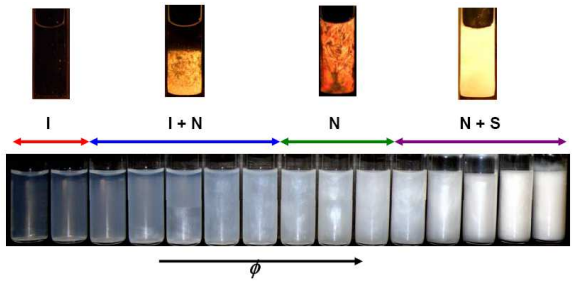}
\caption{\label{Bi}(Color online). Optical image (bottom) and images under cross polarizers 
(top) of samples. Bottom sequence corresponds to platelet 
volume fractions (from left to right) $\phi=0.004$, $0.007$, $0.009$, 
$0.013$, $0.017$, $0.021$,
$0.031$, $0.041$, $0.048$, $0.054$, $0.059$, $0.074$, $0.100$, $0.160$,
and $0.200$. Range of phase stability and phase coexistence
are indicated by double headed arrows.}
\end{figure}
To obtain a more accurate estimation of mean platelet diameter and 
polydispersity, Dynamic Light Scattering (DLS) was performed on the exfoliated ZrP 
monolayer platelets, yielding a platelet diameter of 2295 nm 
(close to the TEM estimation) and a polydispersity 
$\sigma_{\text{\tiny{D}}}=32\%$. 
%
%
To search for possible spatial ordering of the platelets, aqueous suspensions were directly 
characterized by X-ray diffraction (XRD).

\begin{figure}
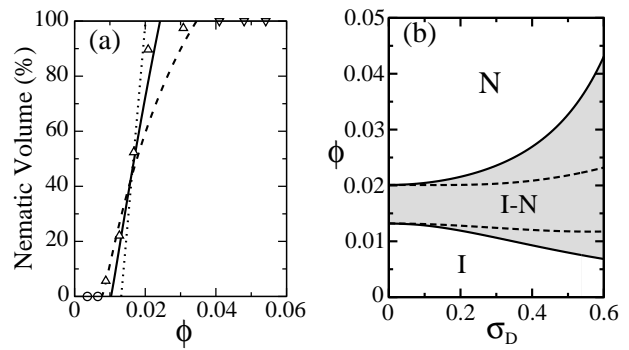

\includegraphics[width=1.6in]{Fig3a.eps}
\includegraphics[width=1.6in]{Fig3b.eps}
\caption{\label{2} (a) Nematic volume fraction (sample volume occupied by the 
N phase) as a function of platelet volume fraction. Symbols: experimental results. Lines: theoretical
results with $\sigma_{\text{\tiny{D}}}=0.00$ (dotted), $0.32$ (continuous), and
$0.52$ (dashed). (b) Theoretical platelet volume fraction as a function of polydispersity
for the I--N transition. Solid lines: I and N cloud curves. Dashed
lines: I and N shadow curves. The coexistence region is shaded.}
\end{figure}

At $\phi\alt 0.01$, the platelet suspension forms a random stable dispersion in water (I phase). 
For samples with high $\phi$, birefringent features through 
crossed polarizers (Fig. \ref{Bi}) 
reveal the occurrence of a transition to a nematic phase for $\phi\agt 0.01$.
In the interval $0.01\alt\phi\alt 0.03$, I and N phases coexist. A full nematic phase is obtained for
$0.03\alt\phi\alt0.06$. The macroscopic phase separation is completed within 24 hours. 
Fig. \ref{2} shows how the fraction of total sample volume occupied by the N phase 
behaves as a function of $\phi$ (symbols). Nonlinearity in the two--phase region is
due to fractionation effects: the diameter polydispersity of the platelets 
in the I and N phases are different (monodisperse suspensions would follow the lever rule and
exhibit linear behavior).

\begin{figure}
\includegraphics[width=2.6in]{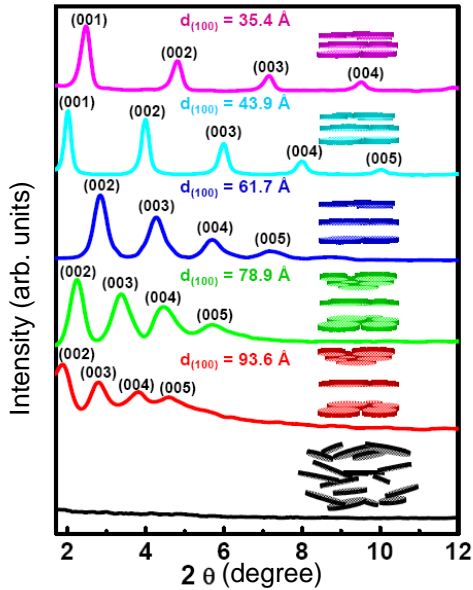}
\caption{\label{Sq} (Color online). XRD intensities as a function of scattering angle
for various volume fractions.
From top to bottom: $\phi=0.200$, $0.160$, $0.100$, $0.074$, $0.059$, and 
$0.054$.}
\end{figure}

\begin{figure}
\includegraphics[width=2.6in]{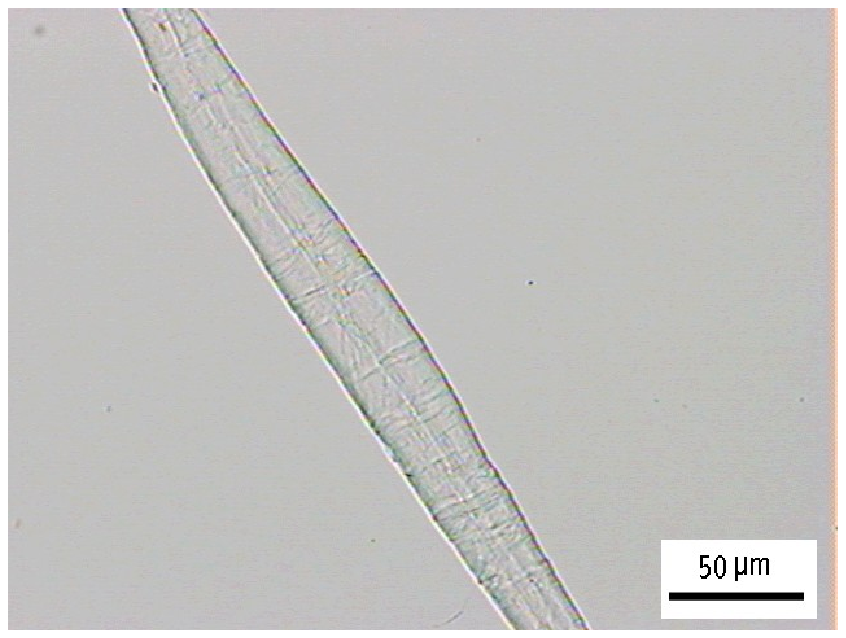}
\caption{\label{ribon} Optical microscopy image of the smectic domains.}
\end{figure}

At high volume fraction of platelets, multiple scattering peaks up to 5 orders in XRD experiments 
are observed (Fig. \ref{Sq}) when $\phi > 0.06$. These scattering peaks are the (001) family of layered arrangements 
of colloidal platelets (S phase). No XRD peaks were obtained for $\phi < 0.06$ (I or 
N phases). The S phase has been 
frequently observed in liquid crystals of rod-like molecules with layered structure. 
However, these organic smectic phases normally lack high-order XRD peaks due to their small domain size. 
For our colloidal platelets, since the size of the mesogen is of the order of hundreds of nanometers to microns, 
the domain size can easily reach tens of microns to millimeters. The presence of higher-order diffraction peaks
and their relatively strong intensity (Fig. \ref{Sq}) is the result of the large domain size. Fig. \ref{ribon} 
shows the optical microscopy image of the smectic liquid crystal observed in our system. A ribbon-like smectic mesophase 
around 50 $\mu$m wide and with a length of up to a few millimeters has been observed. Considering the size of the plate 
mesogens (around 2 $\mu$m, Fig. \ref{1}), the optical microscopy image clearly demonstrates the assembly of 
our colloidal platelets into smectic layers parallel to the 
the plane of the image.

Unfortunately it was not possible to directly measure the nematic, $\phi_N$, and smectic, $\phi_S$,
volume fractions (i.e. fractions of volume occupied by platelets in the respective phase).
An upper limit for $\phi_S$ can be
roughly estimated by assuming smectic layers to consist of single platelet sheets in a random
close--packed arrangement, which gives $\phi_S=\phi_{2D}L/d$, $\phi_{2D}=0.815$ being the
random close--packing area fraction of polydisperse hard disks with
$\sigma_{\text{\tiny{D}}}=0.3$ \cite{Ogata}, and $d$ the smectic period. 
In Table \ref{I} values of $\phi_S$ are collected under
the column labeled {\it max}.

\begin{table}[h]
\caption{\label{I} Characteristics of the smectic phase measured 
in the experiments. $\Delta\phi=\phi-\phi^{*{\rm exp}}$, with $\phi^{*{\rm exp}}\simeq 0.059$
the estimated experimental volume fraction at the spinodal point.}
\begin{ruledtabular}
\begin{tabular}{cccccc}
$\phi$ & $\Delta\phi$ &
$d$ (nm) & $d/L$ & $L_{\rm eff}/L$ &
$\phi_S$ \\
 & & & & & \small{estim. (max.)} \\
\hline
0.059 & 0.000& 9.36 & 3.49 & 2.67 &0.108 (0.233)\\
0.074 & 0.015& 7.89 & 2.94 & 2.27 &0.136 (0.277)\\
0.102 & 0.043& 6.17 & 2.30 & 1.80 &0.193 (0.354)\\
0.161 & 0.102& 4.39 & 1.64 & 1.34 &0.326 (0.497)\\
0.199 & 0.140& 3.54 & 1.32 & 1.11 &0.441 (0.617)\\ 
\end{tabular}
\end{ruledtabular}
\end{table}

In the experiments the S phase occupies a small, slowly expanding, region of the sample.
It seems that the phase transition 
(which may take many days to complete due to the high viscosity of 
the samples) is not proceeding fast enough to complete within the finite observation time. 
The weak diffraction peaks in the XRD spectra for the first smectic state
lend support to a weakly first--order or continuous scenario for the N--S transition. 


One interesting feature is that the layer spacing (Table \ref{I}) exhibits a large
variation (close to three--fold) in the volume--fraction interval explored,
which may point to the existence of long--range platelet repulsion due to platelet 
surface charges or possibly hydration forces (since platelets are hydrophilic, 
water--mediated effects might be at work \cite{Israelichvili}). Attractive van der Waals forces
may also operate at shorter distances.

To rationalize these findings, we map the real particle onto an effective model system 
of parallel polydisperse {\it hard} cylinders of diameter $D$ and thickness $L_{\rm eff}>L$.
Essentially, the model incorporates long--range repulsion via an effective 
particle thickness (a procedure used to model aqueous
suspensions of charged virus particles with added salt \cite{Fraden}). For the moment we simply
note that the value of $L_{\rm eff}$ will depend on $\phi$ and the type of phase.

\begin{figure}
\includegraphics[width=3.in]{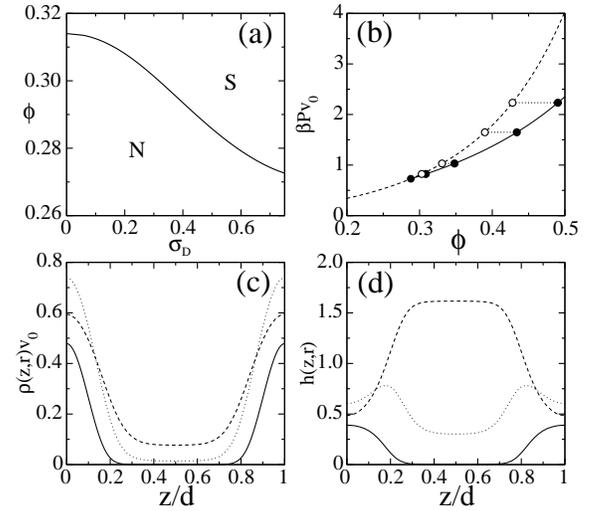}
\caption{
(a) Theoretical volume fraction of the spinodal 
instability against smectic fluctuations as a function of polydispersity.
(b) Osmotic pressure (in reduced units) vs. platelet volume fraction $\phi$ for smectic 
(continuous line) and nematic (dashed line) for $\sigma_{\text{\tiny{D}}}=0.52$. 
Filled circles: smectic states; 
open circles: unstable nematic states with $\phi$ obtained
as $\phi=\phi^{*{\rm the}}+\Delta\phi$, with $\phi^{*{\rm the}}=0.288$ 
(theoretical bifurcation point) 
and $\Delta\phi$ taken from experiment (see Table \ref{I}).
(c) Density profile $\rho(z,r)$ for $r=1.5$
(solid line), $0.4$ (dashed) and $0.8$ (dotted). (d) Normalized density distribution
function $h(z,r)$ in one smectic period. Lines as in panel (c). 
In (c) and (d) the smectic volume fraction is $\phi_S=0.452$ and the
smectic period $d/L_{\rm eff}=1.211$.}
\label{fig4}
\end{figure}

We obtained some predictions for the hard--core model by applying a theoretical analysis based on 
DFT. In the case of the I--N transition we used Onsager theory
\cite{Onsager}, which takes into account excluded--volume effects at second order 
in density and should give accurate results because of the high diameter-to-thickness 
ratio of the platelets.
Since samples are polydisperse, a given distribution of platelet diameters has to be imposed.
Furthermore, the nematic order parameter, measuring the degree of orientation of the particle main axis about 
the director, is expected to be very high and the simplifying Zwanzig approximation 
\cite{Zwanzig} (where only three mutually perpendicular particle orientations, along 
$x$, $y$ and $z$ Cartesian axes, are permitted)
can be implemented. The total free-energy density in reduced thermal units ($k_BT$),
$\Phi=\Phi_{\rm{id}}+\Phi_{\rm{exc}}$, can be split into ideal,
\stepcounter{equation}
\begin{equation*}
\Phi_{\rm{id}}=\int dr\left\{2\rho_{\perp}(r)\left[\ln \rho_{\perp}(r)-1\right]+\rho_{\parallel}(r)
\left[\ln \rho_{\parallel}(r)-1\right]\right\},\nonumber\\ 
\hspace*{-0.25cm}\tag{1}
\end{equation*}
and excess
\stepcounter{equation}
\begin{equation*}
\Phi_{\rm{exc}}(\pi \langle D\rangle^3/4)=2\left[m_{\perp}^{(1)}m_{\perp}^{(2)}+
m_{\parallel}^{(1)}m_{\perp}^{(2)}+m_{\parallel}^{(2)}m_{\perp}^{(1)}\right], \nonumber\\
\hspace*{-0.25cm}\tag{2}
\end{equation*}
parts. 
The moments $m_{\mu}^{(\alpha)}$ of the number density distribution functions 
$\rho_{\mu}(r)$ are defined as 
\begin{eqnarray}
m_{\mu}^{(\alpha)}=\left(\pi \langle D\rangle^3/4\right)\int dr r^{\alpha}\rho_{\mu}(r). 
\end{eqnarray}
We use the dimensionless platelet diameter $r=D/\langle D\rangle$ (with $\langle D\rangle$ 
the mean diameter) as polydispersity variable, and used the notation 
$\rho_x(r)=\rho_y(r)=\rho_{\perp}(r)$ and $\rho_z(r)=\rho_{\parallel}(r)$ to label the densities of 
species oriented perpendicular and parallel to the director, respectively. 
For the I and N symmetries we have 
\begin{eqnarray}
\rho_{\perp}^{\rm{I}}(r)&=&\rho_{\parallel}^{\rm{I}}(r)=\frac{\rho^{\rm{I}}(r)}{3},\\ 
\rho_{\perp}^{\rm{N}}(r)&=&\frac{1-Q(r)}{3}\rho^{\rm{N}}(r),\quad \rho_{\parallel}^{\rm{N}}(r)
=\frac{1+2Q(r)}{3}\rho^{\rm{N}}(r),\nonumber\\
\end{eqnarray}
where $-1/2\leq Q(r)\leq 1$ is the N order parameter of species with reduced diameter 
$r$. 
To find the I and N coexistence densities, i.e. 
\begin{eqnarray}
\rho^{\rm{I,N}}=\sum_{\mu=x,y,z}\int dr \rho_{\mu}^{\rm{I,N}}(r), 
\end{eqnarray}
in a system with a relative fraction of volume occupied by the N phase given by the 
parameter 
$0\leq\gamma\leq 1$, we need to implement the constrained functional minimization 
of 
\begin{eqnarray}
\Phi=(1-\gamma)\Phi^{\rm{I}}+\gamma\Phi^{\rm{N}}, 
\label{total}
\end{eqnarray}
with respect to $\rho_{\mu}^{\rm{I,N}}(r)$.
$\Phi^{I,N}$ are the total 
free-energy densities of the coexisting I and N phases. 
The constraint is dictated by the lever rule restriction: 
\begin{eqnarray}
\rho_0(r)=(1-\gamma) \rho^{\rm{I}}(r)+\gamma\rho^{\rm{N}}(r), 
\end{eqnarray}
where $\rho_0(r)=\rho_0 h(r)$ is the parent number density distribution function with a fixed 
function $h(r)$, which is the diameter distribution function density normalized 
as $\int dr h(r)=1$. We choose a two-parameter function which, after normalising and 
applying the additional condition $\int dr r h(r)=1$ (note that this is equivalent to
$\langle r\rangle=1$), becomes
\stepcounter{equation}
\begin{equation*}
h(r)=\frac{2\Gamma\left[(\nu+2)/2\right]^{\nu+1}}{\Gamma\left[(\nu+1)/2\right]^{\nu+2}}
r^{\nu}\exp\left\{-\left(\frac{\Gamma\left[(\nu+2)/2\right]}{
\Gamma\left[(\nu+1)/2\right]}r\right)^2\right\},\nonumber\\
\hspace*{-0.25cm}\tag{9}
\end{equation*}
where $\Gamma(x)$ is the Gamma function. The free parameter $\nu>0$ controls the degree of polydispersity. 
Note that the present function $h(r)$ has a 
Gaussian tail for $r\gg 1$, while it fulfills $h(0)=0$, in contrast to the Gaussian 
distribution function.
The constrained minimization of Eq. (\ref{total}) and the equality of osmotic pressures
\begin{eqnarray}
\beta P^{\rm{I,N}}=\rho^{\rm{I,N}}+\Phi_{\rm{exc}}^{\rm{I,N}} 
\end{eqnarray}
between both phases allow us to find the coexistence total densities 
as a function of $\gamma$. For 
$\gamma=0$ we find the coexistence between a I phase which occupies the entire volume, 
called the cloud phase, and an infinitesimal amount of the N phase (called shadow phase). 
In the other limit, $\gamma=1$, we find the cloud N--shadow I coexistence.
Results for the I--N transition are plotted in Fig. \ref{fig4}(a) as lines for three 
different polydispersities: $\sigma_{\text{\tiny{D}}}=0.00$ (monodisperse fluid), $0.32$ (same as in experiment), 
and $0.52$. The effective thickness (assumed to be independent of $\phi$) 
was set to $L_{\rm eff}=9L$ so as to make the experimental and theoretical curves with the same value 
of $\sigma_{\text{\tiny{D}}}$ approximately overlap (note however that our theoretical data, without scaling, 
agree well with the experimental results of van der Kooij et al.
\cite{vdKooij} for their approximately hard platelets).
We note that: (i) for high $\sigma_{\text{\tiny{D}}}$ the curvature in the transition region is similar to that 
of the experiment. (ii) The I$+$N coexistence gap is too small in the theory; this may be due to the 
crudeness of the model, and/or to an inaccurate experimental estimation of $\sigma_{\text{\tiny{D}}}$ (note that the 
width of the I$+$N region in the theoretical curve for 
$\sigma_{\text{\tiny{D}}}=0.52$ is quite close to the experimental one). As inferred from Fig.  \ref{2}(b), the coexistence gap 
(defined as the distance between the
I and N cloud curves at the same polydispersity) increases with $\sigma_{\text{\tiny{D}}}$
due to fractionation effects: larger platelets tend to go to the N phase, while smaller platelets 
preferentially populate the I phase.

Next we turn to the theoretical description of the N--S transition. The S phase demands
a description in terms of the density distribution $\rho(z,r)$, where
$z$ is the distance along the nematic director.
From this, the local smectic volume fraction can be calculated,
$\phi_S(z)=v_0\int dr r^2\rho(z,r)$ (with $v_0=\pi\left<D\right>^2L_{\rm eff}/4$ 
an average particle volume), and also a mean
smectic volume fraction, $\phi_S=d^{-1}\int_0^d dz \phi_S(z)$.
A DFT fundamental-measure scheme for strictly parallel hard cylinders \cite{Yuri-Cuesta} 
was generalized to the polydisperse case (the assumption of parallel particles should 
be valid since the nematic order parameter is expected to be very high). The
excess, $z$-dependent local free-energy density in units of $k_BT$ is
\stepcounter{equation}
\begin{equation*}
\Phi_{\rm{exc}}v_0=-n_0\ln(1-n_3)+\frac{2n_1^{\perp}n_2^{\perp}+n_1^{\parallel}
n_2^{\parallel}}{1-n_3}+
\frac{n_2^{\parallel}\left(n_2^{\perp}\right)^2}{\left(1-n_3\right)^2},\nonumber\\
\hspace*{-0.25cm}\tag{11}
\end{equation*}
where the weighted densities are
\begin{eqnarray}
&&n_0(z)={\cal M}^{(0)}\ast \omega^{(0)}(z), 
\quad n_3(z)={\cal M}^{(2)}\ast\omega^{(3)}(z), \nonumber\\
&&n_1^{\perp}(z)={\cal M}^{(1)}\ast 
\omega^{(0)}(z), \quad n_1^{\parallel}(z)={\cal M}^{(0)}\ast\omega^{(3)}(z),\nonumber \\ 
&&n_2^{\perp}(z)={\cal M}^{(1)}\ast\omega^{(3)}(z), \quad 
n_2^{\parallel}(z)={\cal M}^{(2)}\ast\omega^{(0)}(z),\nonumber\\ 
\label{nes}
\end{eqnarray}
with
${\cal M}^{(\alpha)}(z)=m^{(\alpha)}_{\parallel}(z)/\kappa$ and
$\kappa=\langle D\rangle/L_{\rm{eff}}$. The weights are
\begin{eqnarray}
\omega^{(0)}(z)=\frac{1}{2}\delta\left(\frac{1}{2}-|z|\right), \quad \omega^{(3)}(z)=
\Theta\left(\frac{1}{2}-|z|\right),\nonumber\\ 
\label{omegas}
\end{eqnarray}
with $\delta(z)$ and $\Theta(z)$ the usual Dirac-delta 
and Heaviside-step functions; the symbol $\ast$ denotes convolution, i.e. 
\begin{eqnarray}
f\ast\omega^{(0)}(z)&=&\frac{1}{2}\left[f(z-1/2)+f(z+1/2)\right], \nonumber\\ 
f\ast\omega^{(3)}(z)&=&\int_{z-1/2}^{z+1/2}dz'f(z').
\label{conv}
\end{eqnarray}
In all preeceding equations (\ref{nes})--(\ref{conv}) the coordinate $z$ is in units of
$L_{\rm{eff}}$.
The  
ideal and the interaction free-energy functionals per unit volume  
for the smectic are then
\begin{eqnarray}
\hspace*{-0.1cm}
&&\beta {\cal F}_{\rm{id}}/V=d^{-1}\int_0^ddz\int dr\rho(z,r)\left[\ln \rho(z,r)-1\right],\nonumber\\
&&\\
&&\beta {\cal F}_{\rm{exc}}/V=d^{-1}\int_0^ddz\Phi_{\rm{exc}}(z). 
\end{eqnarray}
The equation of state of S phase was obtained by functional minimization of 
the total free-energy with respect to $\rho(z,r)\equiv\rho_{\parallel}(z,r)$ with 
the restriction $\rho_0(r)=d^{-1}\int_0^{d} dz\rho(z,r)$, $\rho_0(r)$ being the 
(fixed) mean density distribution function per period.
The N--S transition is found to be continuous for all values of polydispersity (see Fig. \ref{fig4} (a),
where the limit $\sigma_{\text{\tiny{D}}}=0$ corresponds to the one--component model of Ref. \cite{Yuri-Cuesta1}). 
This is in qualitative agreement with experiments which, as discussed above, suggest a continuous or 
weakly first--order phase transition \cite{remark0}. 

Fig. \ref{fig4}(b) shows the osmotic pressure for N and S phases as a function
of $\phi$ for the hard--core potential. The N phase is unstable beyond the bifurcation point.
Even though there are no signs of N$+$S coexistence or macroscopic fractionation at 
the transition, some subtle micro--fractionation effects in the S phase exist, shown 
in Figs. \ref{fig4}(c-d) which pertain to values $\sigma_{\text{\tiny{D}}}=0.52$ and $\phi_S=0.452$. 
Density profiles for  
different diameters show micro--fractionation: the probability density
$h(z,r)=\rho(z,r)/\int dr \rho(z,r)$ shows that the 
relative fraction of large platelets is higher at smectic layers, while
the opposite occurs for smaller platelets, with maxima at interstitials. The relative 
fraction of intermediately--sized platelets has two maxima per period, located a bit off the layers.

Finally we extract some information from the model in order to interpret experimental data and,
in particular, to refine the values for $\phi_S$ previously derived as upper estimates.
We assume the experimental transition to occur at a 
platelet volume fraction $\phi^*\approx 0.059$ (interpreted as a spinodal point or as the point where the
smectic phase appears for the first time). The experimentally 
observed N states for $\phi\agt\phi^*$ would then be unstable, with nematic volume fractions $\phi_N<\phi_S$; 
here $\phi_S$ is the (unknown) platelet volume fraction of the stable smectic state (at the 
same osmotic pressure). Since the smectic volume in the samples is small, we take $\phi_N\simeq\phi$. 
Also, we identify the (unstable) experimental nematic states with the theoretical (unstable) nematic 
branch. We predict $\phi_S$ from our equation of state, Fig. \ref{fig4}(b), as follows. First, the 
volume--fraction distance $\Delta\phi=\phi-\phi^*$ is taken to be the same in both theory and experiment, 
which allows to obtain, for each $\phi$ (or $\Delta\phi$) in Table \ref{I}, a theoretical value
$\phi_S^{\rm the}$ [indicated by filled circles in Fig. \ref{fig4}(b)], to which our
hard--core theory uniquely associates a value $d/L_{\rm eff}$, and estimate $L_{\rm eff}/L=(d/L)/(d/L_{\rm eff})$
(see Table \ref{I}); from low to high volume fraction, $L_{\rm eff}/L=2.67$--$1.11$. 
Now the estimated $\phi_S$ for the real system follows from $\phi_S=\phi_S^{\rm the}/(L_{\rm eff}/L)$ --see
Table \ref{I} under column {\it estim.} \cite{remark}.
Even though the refined procedure probably gives more reasonable values for 
$\phi_S$ than the previous crude estimates, we cannot at present improve the model for lack of
accurate experimental measurements. However, all evidence points to the existence of a weak phase transition 
from a N to a {\it stable} S phase in this novel colloidal suspension of platelets.

To summarize, we have experimentally analyzed a new highly polydisperse colloidal platelet suspension exhibiting 
smectic ordering, thus confirming that this elusive phase can be stabilized in platelet colloids when 
polydispersity suppresses the columnar phase. A theoretical model for polydisperse platelets based on DFT 
and an effective hard--core model partially explains the experimentally observed phase behavior.

We thank J. A. Cuesta and A. Clearfield for useful discussions and a critical 
reading of the manuscript. Acknowledgment is made to the donors of ACS Petroleum Research 
Fund (45303-G7) 
and to the Dow Chemical Company. This work has been partly financed by start-up funds from Texas 
Engineering Experimental Station and Texas A\&M University, by grants 
NANOFLUID, MOSAICO and S-0505/ESP-0299 from Comunidad Aut\'onoma de 
Madrid (Spain), and Grants FIS2005-05243-C02-01, FIS2007-65869-C03-01,
FIS2008-05865-C02-02 and FIS2007-65869-C03-C01 from 
Ministerio de Educaci\'on y Ciencia (Spain).


\begin{references}
\bibitem{remark0} Further theoretical calculations with attractive forces (not shown) indicate that the transition
changes to first order when the system is polydisperse (orientational
degrees of freedom might have some additional effect on the transition).
\bibitem{remark} Note that the smectic volume fraction $\phi_S^{\rm the}$
and period $d/L_{\rm eff}$ at the spinodal point could both decrease due to additional attractive, short-ranged
forces plus long--ranged, repulsive forces; in this case 
$L_{\rm eff}/L$ would increase and consequently $\phi_S$ would be lower than our estimate ($0.108$) using 
a purely repulsive potential.
\end{references}
\end{document}